\documentclass[10pt,letterpaper]{article}
\usepackage{opex3}
\usepackage{cite}
\usepackage{graphicx}

\begin{document}

\title{Increased mode instability thresholds of fiber amplifiers by gain saturation}

\author{Arlee V. Smith$^*$ and Jesse J. Smith}

\address{AS-Photonics, LLC, 6916 Montgomery Blvd. NE, Suite B8, Albuquerque, NM 87109 USA}

\email{$^*$arlee.smith@as-photonics.com}

\begin{abstract}
We show by numerical modeling that saturation of the population inversion reduces the stimulated thermal Rayleigh gain relative to the laser gain in large mode area fiber amplifiers. We show how to exploit this effect to raise mode instability thresholds by a substantial factor. We also demonstrate that when suppression of stimulated Brillouin scattering and the population saturation effect are both taken into account, counter-pumped amplifiers have higher mode instability thresholds than co-pumped amplifiers for fully Yb$^{3+}$ doped cores, and confined doping can further raise the thresholds.
\end{abstract}


\section{Introduction}

In earlier papers we described a stimulated thermal Rayleigh scattering process (STRS) that can account for observed modal instability in large mode area fiber amplifiers, and we described in detail our numerical model of this process\cite{smithsmith2011,smithsmith2013,smithsmith2012a,smithsmith2013b,smithsmith2013c}. Our model computes the mode coupling gain and the laser gain for fiber amplifiers with practically achievable step index profiles operating at realistic pump and signal powers. It was shown that the quantum defect heating associated with laser amplification, in conjunction with a frequency offset between the strong fundamental mode and the weak parasitic higher order mode, leads to a high exponential gain for the parasitic mode. Consequently, mode instability has a sharp power threshold, above which the output beam quality is severely degraded. 

The essence of the STRS process responsible for mode instability is that laser gain necessarily deposits quantum defect heat in the core of the amplifier fiber, and the asymmetric heating produced by the asymmetric signal irradiance profile due to interference between modes LP$_{01}$ and LP$_{11}$ leads to an asymmetric thermal lens that couples light between those two modes. An additional requirement is that there be a phase shift between the temperature grating and the signal irradiance grating which, in our STRS model, is created by the time lag between an irradiance grating traveling along the fiber and the temperature grating that it creates.

Our model imposes a steady-periodic condition on the temperature grating because we assume the frequency offset between modes has a narrow linewidth. Alternative STRS models have been presented by Hansen {\it et al.}\cite{hansen,hansen2013} and by Dong\cite{dong}. They used similar approximations, including the steady-periodic assumption, and they predict instability thresholds similar to ours. Another model by Ward {\it et al.}\cite{ward} is based on STRS, but without the steady-periodic assumption. It also predicts thresholds similar to ours.

The primary difference in the physics of the various models is that the models of Hansen {\it et al.} and Dong assume the profile of the quantum defect heating matches the profile of the signal light, while our model and that of Ward {\it et al.} compute the heat profile based on the local change in either the pump or signal irradiance calculated using the local upper state population. In the model of Ward {\it et al.} the increase of signal power in a mode due to laser gain is found from the overlap of the local gain $g(x,y,z)$ with the field of that mode. In our BPM model laser gain increases the total signal field locally and is then apportioned among the modes automatically by diffraction in the presence of the core index step. Either the local signal field growth or the local pump irradiance loss is used to compute the quantum defect heating.

Hansen {\it et al.} and Dong showed that using their assumed heat profile the mode coupling gain is related in a simple way to the laser gain. This would imply that the mode instability threshold is largely determined by the net laser gain, and can be adjusted only by changing the modal profiles and their overlap with the Yb$^{3+}$ doping profile. In fact, they show that by using a low value of the $V$ parameter or by confining the Yb$^{3+}$ doping to the central portion of the core, the threshold is raised. However, we show in this report that the existence of depletion of the upper state Yb$^{3+}$ population breaks the simple connection between laser gain and mode coupling gain, making it possible to design fibers with substantially higher mode coupling thresholds than predicted by Hansen {\it et al.} and by Dong. An amplifier that is designed to reach a specific level of laser amplification can be designed to avoid the mode instability. Specifically, by designing the amplifier so it has a higher ratio of pump cladding diameter to core diameter, the STRS threshold can be raised. Of course this requires a longer fiber and a smaller core, a combination that is problematic for SBS suppression. In practice the amplifier must achieve a suitable balance between STRS and SBS suppression. We will also discuss how to achieve this balance. 

\section{Transverse hole burning}

In computing the upper state population fraction $n_u$ we use the steady state expression
\begin{equation}\label{eq.nuss}
n_u(x,y)=\frac{I_p\sigma_p^a/h\nu_p+I_s(x,y)\sigma_s^a/h\nu_s}{I_p(\sigma_p^a+\sigma_p^e)/h\nu_p+I_s(x,y)(\sigma_s^a+\sigma_s^e)/h\nu_s+1/\tau}.
\end{equation}
Here the $\sigma$'s are the absorption and emission cross sections for the pump and the signal, the $\nu$'s are the optical frequencies, and $I_p$ and $I_s$ are the pump and signal irradiances. The amplifier parameters used throughout this report are listed in Table 1. We model a fiber with a 50 $\mu$m diameter step index core with a typical numerical aperture of 0.054. The Yb$^{3+}$ doping density is also typical of high power amplifiers.
\begin{table}[htb]
\centering\caption{Amplifier parameters}
\begin{tabular}{cc|cc}\hline 
$d_{core}$&50 $\mu$m&$d_{dope}$ &30-50 $\mu$m\\
$d_{clad}$&100-500 $\mu$m&$N_{Yb}$&3.0$\times 10^{25}$ m$^{-3}$\\
$\lambda_p$&976 nm&$\lambda_s$&1032 nm\\
$\sigma_p^a$&2.47$\times 10^{-24}$ m$^2$&$\sigma_p^e$&2.44$\times 10^{-24}$ m$^2$\\
$\sigma_s^a$&5.80$\times 10^{-27}$ m$^2$&$\sigma_s^e$&5.0$\times 10^{-25}$ m$^2$\\
${P}_p$&varies&${P}_s$&10 W\\
$dn/dT$&1.2$\times 10^{-5}$&$L$&varies\\
$\rho$&2201 kg/m$^3$&$C$&702 J/kg$\cdot$K\\
$n_{core}$&1.451&$n_{clad}$&1.45\\
$\tau$&901 $\mu$s&$K$&1.38 W/m$\cdot$K\\
$NA$&0.054&$V$&8.2\\
$A_{\rm eff}\;(\mbox{LP$_{01}$})$& 1175 $\mu$m$^2$&&\\ 
\hline
\end{tabular}
\end{table}

\begin{figure}[htbp]
\centering
\includegraphics[width=5.25in]{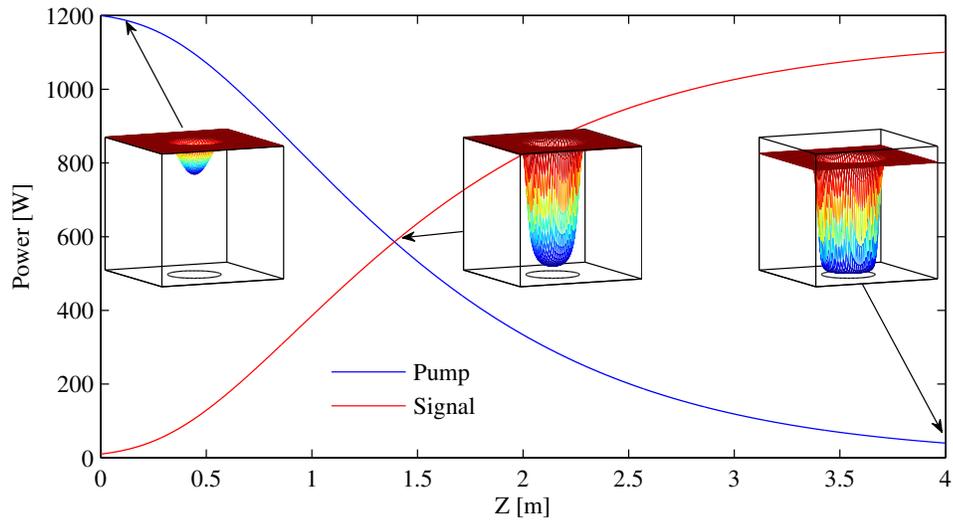}
\caption{\label{fig.saturation_co}Signal and pump powers versus $z$ for co-pumped fiber with $d_{\rm core}=d_{\rm dope}=50$ $\mu$m, $d_{\rm clad}=400$ $\mu$m, operating at the mode instability threshold. The inset figures show the upper state fraction $n_u$ profiles at $z=0.1$, 1.4, and 4.0 m. The range of the $n_u$ axis is 0-0.5. The circles on the bottom faces of the frames indicate the edge of the core.}
\end{figure}
\begin{figure}[htbp]
\centering
\includegraphics[width=5.25in]{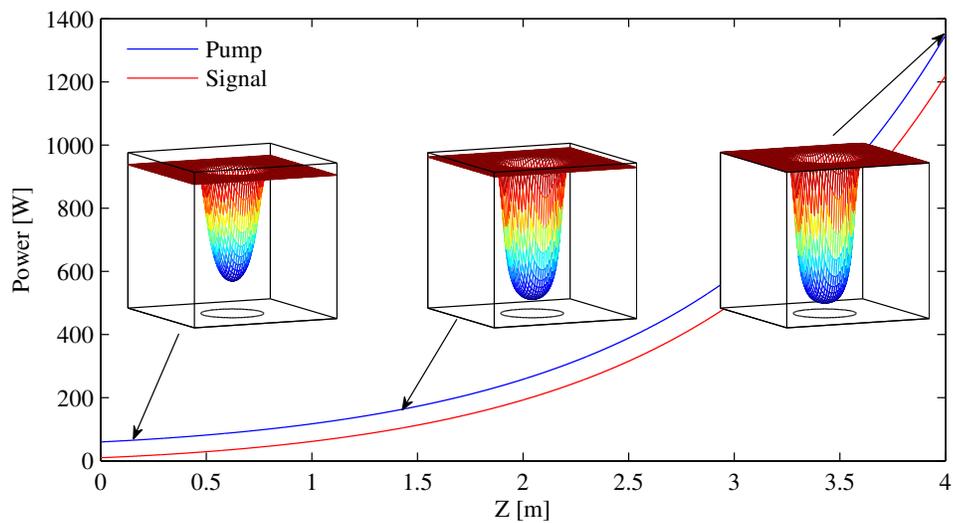}
\caption{\label{fig.saturation_counter}Same fiber and $z$ locations as in Fig. \ref{fig.saturation_co} except the amplifier is counter-pumped instead of co-pumped. The threshold powers are slightly different for co- and counter-pumped cases.}
\end{figure}
Figures \ref{fig.saturation_co} and \ref{fig.saturation_counter} show upper state population profiles, $n_u(x,y)$ computed at three points along co-pumped and counter-pumped fiber amplifiers with a cladding diameter of 400 $\mu$m. The signal light is all in LP$_{01}$ here. Near the input end of the co-pumped fiber the pump is strong and the signal is weak so the upper state population is weakly depleted. The shape of the population depletion closely matches the signal irradiance profile here. The undepleted population in the region with $I_s=0$ is $n_u=\sigma_p^a/(\sigma_p^a+\sigma_p^e)\approx 0.5$. Further along this fiber, near the crossing of the signal and pump powers, the saturation is strong at the mode center but moderate near the core boundary, while near the output end saturation is strong across the entire core. If the same fiber is counter-pumped the signal and pump powers are approximately equal along the full length of the fiber, so the saturation resembles that of the middle diagram of the co-pumped fiber where the signal and pump powers are equal. The degree of saturation would be reduced in a fiber with a smaller cladding, and increased in one with a larger cladding. 

The quantum defect heating is proportional to the pump absorption, which is determined from the value of $n_u(x,y)$ and the pump irradiance. The latter is assumed uniform across the pump cladding. We compute the heat deposition rate $Q$ using
\begin{equation}
Q(x,y)=N_{Yb}(x,y)\biggl[\frac{\nu_p-\nu_s}{\nu_p}\biggr]\biggl[\sigma_p^a-(\sigma_p^a+\sigma_p^e)n_u(x,y)\biggr]{I_p},
\end{equation}
where the first term in brackets is the quantum defect. The heat profile can be visualized by inverting the depletion profiles in Figs. \ref{fig.saturation_co} and \ref{fig.saturation_counter} and multiplying them by the doping profile $N_{Yb}(x,y)$.

\begin{figure}[htbp]
\centering
\includegraphics[width=4.5in]{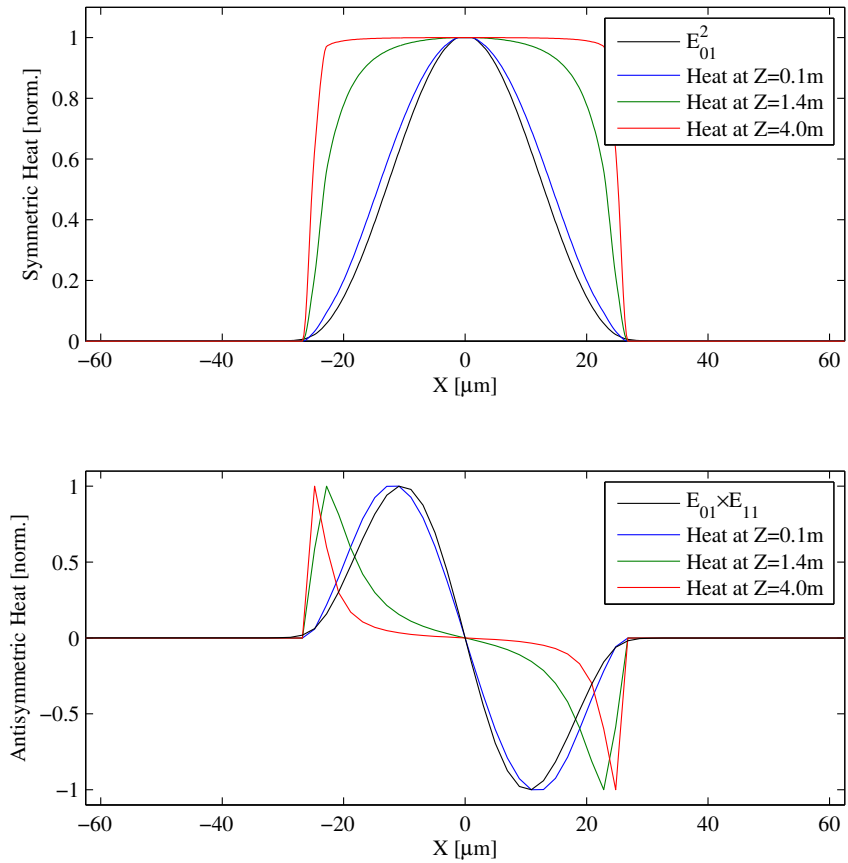}
\caption{\label{fig.heats1}Normalized symmetric part of the heat profile (upper plot) and antisymmetric oscillatory portion of the heat profile (lower plot) for same co-pumped fiber at the same three $z$ locations indicated in Fig. \ref{fig.saturation_co}. Near the input end the symmetric heat profile closely resembles the LP$_{01}$ irradiance profile, while the oscillatory part resembles the product of the fields of LP$_{01}$ and LP$_{11}$. Farther along the fiber the symmetric part of the heat profile becomes nearly flat topped while the antisymmetric part is strongly suppressed near the center ($x=0$).}
\end{figure}

\begin{figure}[htbp]
\centering
\includegraphics[width=4.5in]{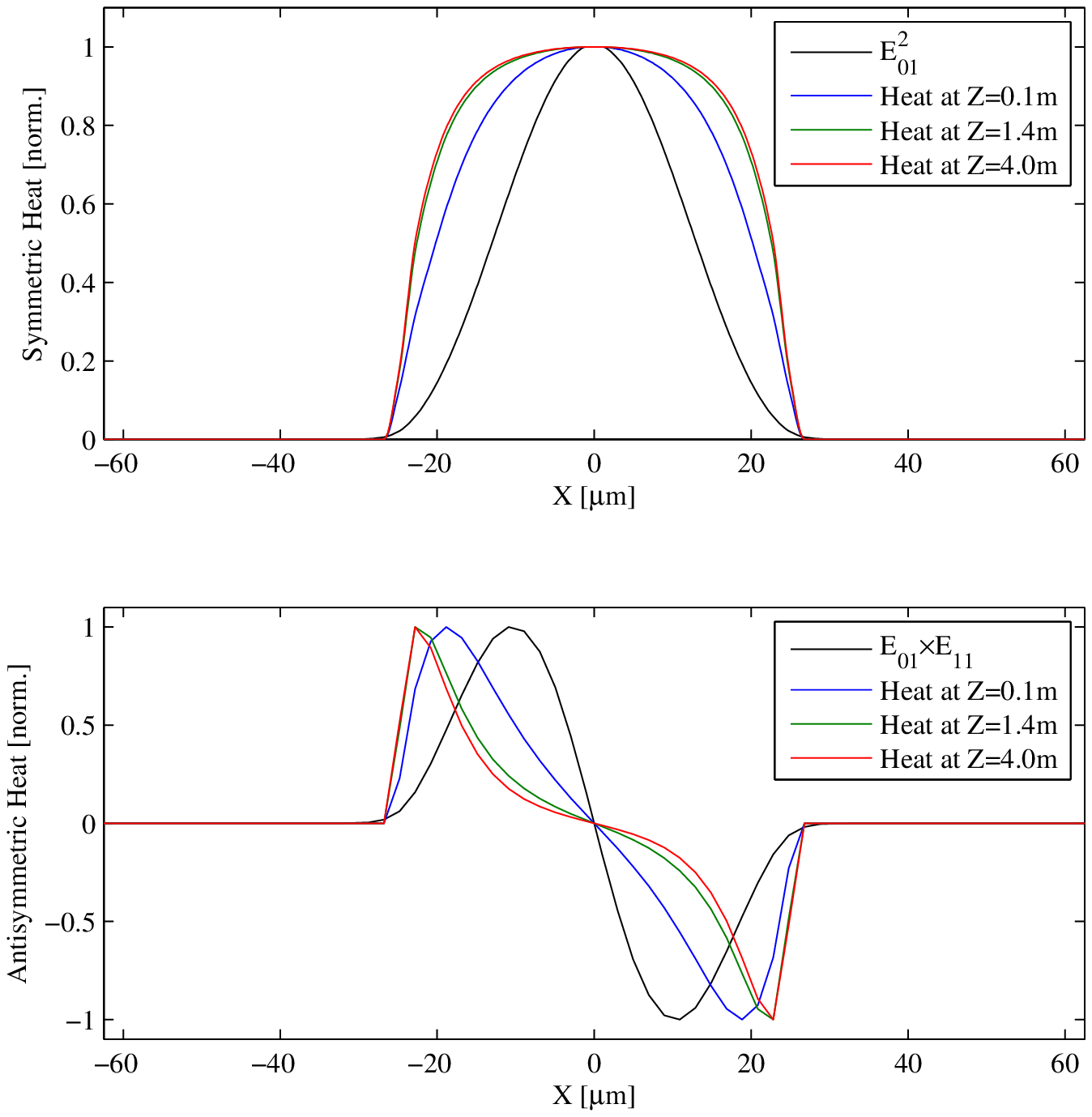}
\caption{\label{fig.heats2}Normalized symmetric part of the heat profile (upper plot) and antisymmetric oscillatory portion of the heat profile (lower plot) for the same counter-pumped fiber at the same three $z$ locations indicated in Fig. \ref{fig.saturation_counter}. The oscillatory anti symmetric heat profile never closely matches the product of the fields of LP$_{01}$ and LP$_{11}$.}
\end{figure}

The portion of the heat profile that is responsible for mode coupling gain (STRS gain) is the antisymmetric part created by the antisymmetric part of the signal irradiance. The population saturation strongly influences the shape of this part of the heating. In Figs. \ref{fig.heats1} and \ref{fig.heats2} we show core centered cuts through the oscillatory heat profiles at the same three $z$ locations as in Figs. \ref{fig.saturation_co} and \ref{fig.saturation_counter}. These heat profiles create the oscillating asymmetric part of the temperature profile that couples modes LP$_{01}$ and LP$_{11}$ in the STRS process. From an examination of these heat profiles it will come as no surprise when we show in the next section that strong saturation causes a significant reduction in STRS gain.

\section{Mode coupling gain}

\subsection{Without hole burning}

Both Dong and Hansen {\it et al.} showed that if the heat profile matches the signal irradiance profile over the portion of the core that is (uniformly) doped, the gain of mode LP$_{11}$ satisfies
\begin{equation}\label{eq.DandH}
\frac{\partial P_{11}(z)}{\partial z}=\Bigl[g_{11}+g_{01}\;\chi\;P_{01}(z)\Bigr]\;P_{11}(z)=g_{\rm net}\;P_{11}(z).
\end{equation}
Here we have simplified Dong's expression by assuming there is no depletion of mode LP$_{01}$ due to STRS, and no linear loss for either mode. The total gain $g_{\rm net}$ for LP$_{11}$ is its laser gain, indicated by $g_{11}$, plus the STRS gain, indicated by $(g_{01}\chi P_{01})$, where $g_{01}$ is the laser gain for LP$_{01}$. The quantum defect heat deposited by laser amplification of LP$_{01}$ is $(g_{01}P_{01})$ so $\chi$ is a real valued coefficient that relates quantum defect heating to STRS gain. For each fiber design $\chi$ has a constant value determined by the frequency offset between LP$_{01}$ and LP$_{11}$ and the quantum defect, plus the thermal, geometrical, and optical properties of the fiber.  

Under the assumption that the shape of the heat profile matches the irradiance profile, $\chi$ is independent of $z$ and the modal powers. It depends on the spatial overlap of the two modes with one another and with the dopant profile. Hansen {\it et al.} and Dong both showed that confined doping reduces the value of $\chi$, as does reducing the $V$ parameter below 5 or so.

For full doping and for $V\ge 5$, Hansen {\it et al.} and Dong both showed that, according to their models, if the starting noise level in LP$_{11}$ is set to approximately 10$^{-16}$ W corresponding to quantum noise, the threshold power lies near 400 W. Details of the fiber design other than $V$ and the doping profile do not affect this threshold. Using a starting power of 10$^{-8}$ W rather than 10$^{-16}$ W reduces the threshold power to approximately 200 W. Confining the doping diameter to 50\% of the core diameter was found to approximately double the threshold power. 

We used our model to verify the value of $\chi$ for one case analyzed by Hansen {\it et al.}(Fig. 4 of ref. \cite{hansen2013}). In order to avoid population saturation we used a pump power of 20 kW, $d_{\rm core}=40$ $\mu$m, $d_{\rm clad}=250$ $\mu$m, and an LP$_{01}$ seed power of 50 W. Otherwise the fiber parameters were those from Table 1. Our frequency of maximum gain was equal to that of Hansen {\it et al.}, and our computed a value for $\chi$ was within 5\% of that of Hansen {\it et al.} Of course, our fiber was extremely inefficient, using 20 kW to produce approximately 400 W of signal, and depleting the pump by only 2.5\%. However, this exercise does demonstrate good agreement between our model and those of Hansen {\it et al.} and Dong in the limit of low saturation.

\subsection{With hole burning}

\begin{figure}[htbp]
\centering
\includegraphics[width=4.5in]{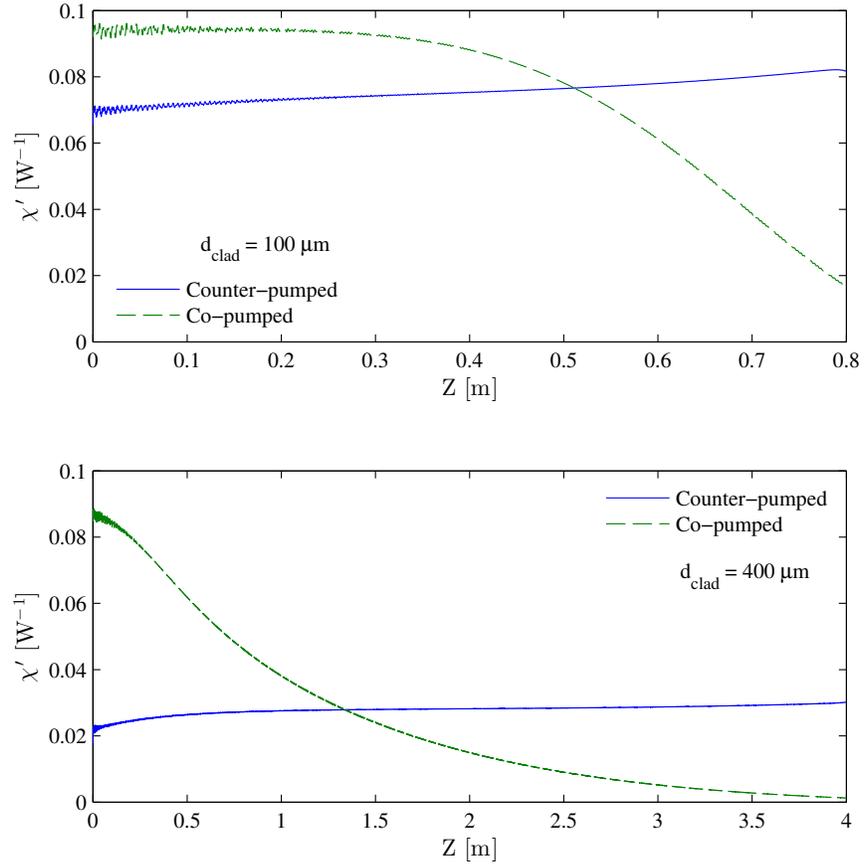}
\caption{\label{fig.chi_vs_saturation}Plots of $\chi^{\prime}$ versus $z$ for co-pumped (dashed green curve) and counter-pumped (solid blue curve) fibers operating near the mode instability threshold. The fiber parameters: 50 $\mu$m diameter core and doping, 100 $\mu$m diameter pump cladding (upper plot) and 400 $\mu$m diameter pump cladding (lower plot), 1100 Hz red detuning of LP$_{11}$, $\lambda_s=1032$ nm, $\lambda_p=976$ nm, $NA$=0.054 ($n_{\rm core}=1.451$, $n_{\rm clad}=1.45$). In the upper plot the pump powers are 525 W co-pumped, and 493 W counter-pumped. In the lower plot the pump powers are 1200 W co-pumped, and 1350 W counter-pumped.}
\end{figure}

The model comparison just described illustrates the lack of realism of the assumption of equal shapes for the heat and irradiance profiles. For the heat profile to match the light profile the pump irradiance must be much stronger than the signal irradiance. However, amplifiers that efficiently convert pump power necessarily experience strong population saturation. To emphasize this contrast between our model and one that assumes matching heat and irradiance profiles and thus constant $\chi$ values, we define a new shape factor $\chi^{\prime}$ as
\begin{equation}
\chi^{\prime}=\frac{g_{\rm comp}-g_s}{g_s P_s}=\frac{g_{\rm strs}}{g_s P_s}
\end{equation}
where $g_{\rm comp}$ is the total gain of LP$_{11}$ computed with our model, replacing the $g_{\rm net}$ of Eq. \ref{eq.DandH}, $g_s$ is the computed signal laser gain, and $g_{\rm strs}$ is the STRS gain. We do not distinguish between the nearly equal laser gains for the two modes in this high $V$ fiber with fully doped core. Variables $P_s$, $g_s$, and $g_{\rm strs}$ are all $z$ dependent computed values. The denominator is proportional to the total deposited heat. Figure \ref{fig.chi_vs_saturation} shows plots of $\chi^{\prime}$ for co-pumped and counter-pumped amplifiers. The upper plot is for a 100 $\mu$m diameter cladding; the lower plot is for a 400 $\mu$m diameter cladding. Saturation effects are stronger with the larger cladding because of the reduced pump irradiances. In both fibers the value of $\chi^{\prime}$ near $z=0$ for the co-pumped case is nearly equal to the $\chi$ of Dong and of Hansen {\it et al.}, as expected because of the low degree population saturation there. As $z$ increases the population saturation strengthens so the value of $\chi^{\prime}$ falls, implying the mode coupling gain is reduced relative to the laser gain. This gain reduction raises the STRS threshold relative to that of Hansen {\it et al.} and Dong. In the small cladding fiber the reduction does not occur until half way along the fiber while in the large cladding fiber it occurs much sooner. 

As expected from the nearly constant ratio of pump to signal for the counter-pumped fibers, the value of $\chi^{\prime}$ is nearly constant along the fiber. However, its value is still reduced to a level seen in the co-pumped fiber near the crossing point near $z=0.5$ m for the small clad fiber and near $z=1.4$ m for the large clad fiber. It is clear from a comparison of the values of $\chi^{\prime}$ that the STRS threshold should be higher for larger cladding sizes, and the threshold for an efficient amplifier should exceed the Hansen {\it et al.} and Dong threshold.

\begin{figure}[tbhp]
\centering
\includegraphics[width=4.5in]{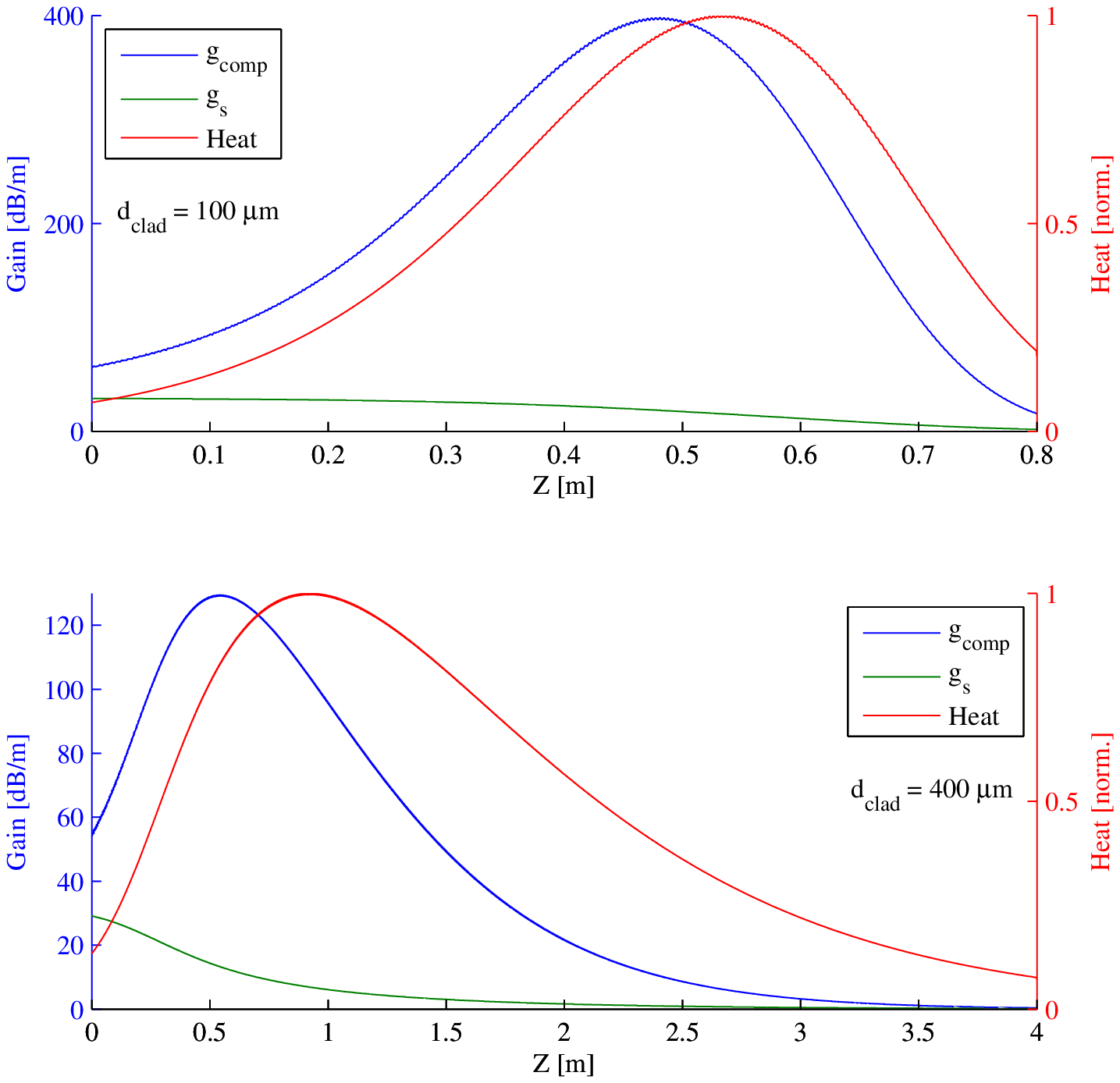}
\caption{\label{fig.ggh}Plots of laser gain $g_s$, total gain $g_{\rm comp}$ and heat versus $z$ for the same co-pumped fibers and the same operating conditions as in Fig. \ref{fig.chi_vs_saturation}. The upper plot is for a 100 $\mu$m diameter cladding; the lower plot is for a 400 $\mu$m diameter cladding. In both plots it is clear that the gain and heat profiles are not closely matched.}
\end{figure}

Another way to view the same information is to plot $g_{\rm comp}$, $g_s$, and the total heat rather than $\chi^{\prime}$. This is done in Fig. \ref{fig.ggh} for the same pair of fibers, co-pumped, with $d_{\rm clad}=100$ $\mu$m (upper plot) and 400 $\mu$m (lower plot).

\begin{figure}[htbp]
\centering
\includegraphics[width=5in]{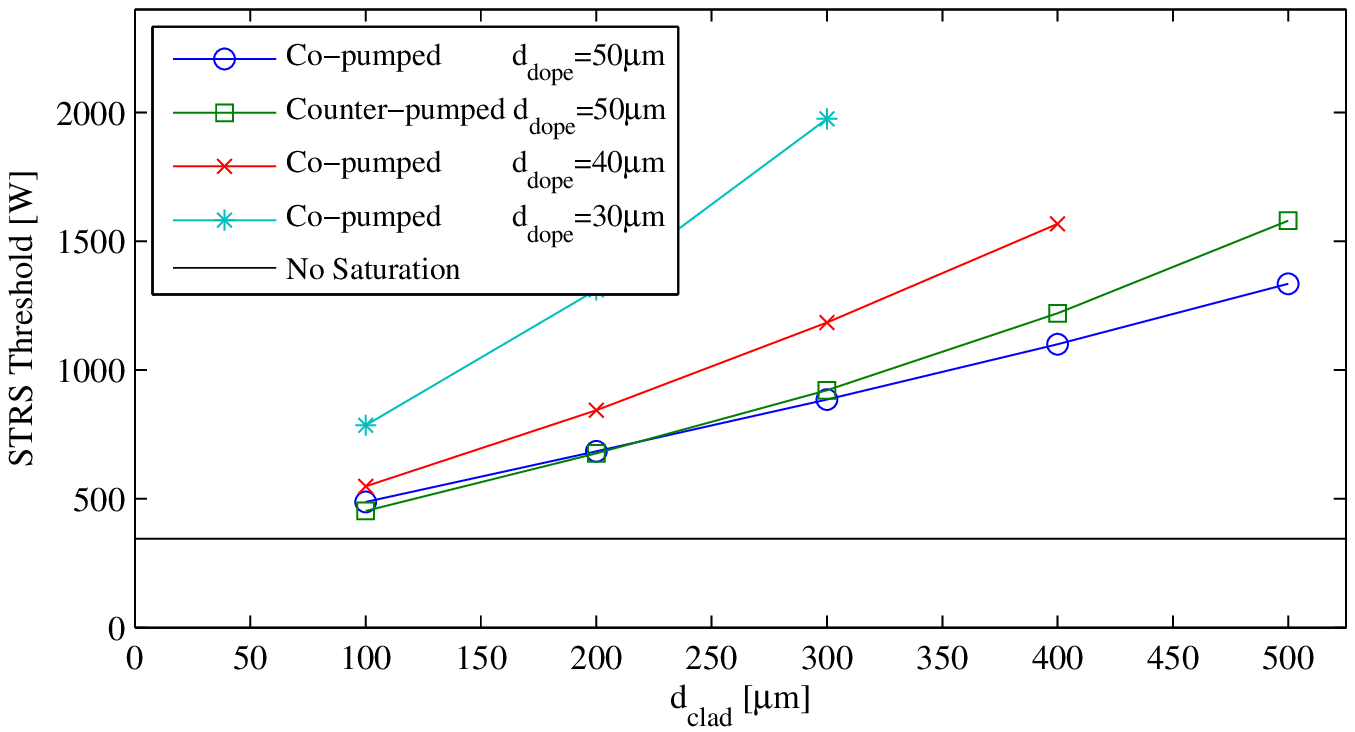}
\caption{\label{fig.thresholds}Output signal powers at the instability threshold for different pump cladding diameters (100, 200, 300, 400, 500 $\mu$m). The parameters: $d_{\rm core}=50$ $\mu$m, $\lambda_p=976$ nm, $\lambda_s=1032$ nm, signal seed powers are 10 W in LP$_{01}$ and $10^{-16}$ W in LP$_{11}$. The lengths of the fibers are adjusted the minimum value necessary to achieve high efficiency, defined as pump absorption $>0.95$. The solid curve at 345 W indicates the threshold computed in the limit of no population saturation. Details are given in the tables of Section \ref{sec.Tables}.}
\end{figure}

The expected trend of higher thresholds for larger cladding sizes is illustrated in Fig. \ref{fig.thresholds}. The computed thresholds, defined as 1\% of the signal power in LP$_{11}$, are seen to rise with increasing cladding diameters and the resulting increasing degree of population saturation. Co- and counter-pumped fibers are found to have similar thresholds. Decreasing the doping diameter also increases the threshold because saturation is then strong across the full doping profile. For comparison we show as a black horizontal line in the figure the threshold predicted without saturation. More details for the model runs included in Fig. \ref{fig.thresholds} are listed in Tables 2-5.

\section{Stimulated Brillouin scattering (SBS) suppression}

Designing a narrow bandwidth fiber amplifier suitable for beam combining applications requires balancing SBS and STRS. Here we present a simplified version of this balancing act. We define an effective SBS gain length at the STRS threshold by
\begin{equation}
P_{\rm thres}L_{\rm eff}=\int_0^L P_s(z) dz.
\end{equation}
The SBS threshold power\cite{agrawal} is exceeded if
\begin{equation}
\frac{g_B}{\gamma}\frac{P_{\rm thres}L_{\rm eff}}{A_{\rm eff}}> 17,
\end{equation}
where the 17 comes from the usual SBS threshold gain value of 21 minus the laser gain for the Stokes wave\cite{agrawal}. The $\gamma$ parameter is an SBS gain reduction factor that reduces the effective SBS gain from its nominal value of $g_B=5\times 10^{-11}$ m/W for silica. The linewidth for silica is approximately 50 MHz and the Stokes shift is approximately 16 GHz for $\lambda=1032$ nm. The value of $\gamma$ can be increased by broadening the signal linewidth by phase modulation\cite{agrawal} above 50 MHz, for example, or by introducing a temperature gradient\cite{hildebrandt} or a strain gradient\cite{horiguchi,rothenberg} along the fiber length to vary the Stokes shift. Another approach is to vary the acoustic velocity, and thus the Brillouin shift, across the core region\cite{li,robin}. Temperature and strain shift the SBS frequency by known amounts, so from the signal linewidth and the temperature and strain gradient one can estimate $\gamma$ using an SBS model. To avoid SBS the value of $\gamma$ must satisfy
\begin{equation}
\gamma>\frac{g_B P_{\rm thres}L_{\rm eff}}{17\;A_{\rm eff}}.
\end{equation}
For our 50 $\mu$m diameter core, $A_{\rm eff}=1175$ $\mu$m$^2$, so the SBS threshold condition is
\begin{equation}\label{eq.gamma}
\gamma=\frac{P_{\rm thres}L_{\rm eff}}{400\; {\rm W}\cdot{\rm m}}.
\end{equation}
If $P_{\rm thres}L_{\rm eff}>400$ W$\cdot$m, this expression gives the minimum value of $\gamma$ necessary to suppress SBS.

\begin{figure}[htbp]
\centering
\includegraphics[width=4.5in]{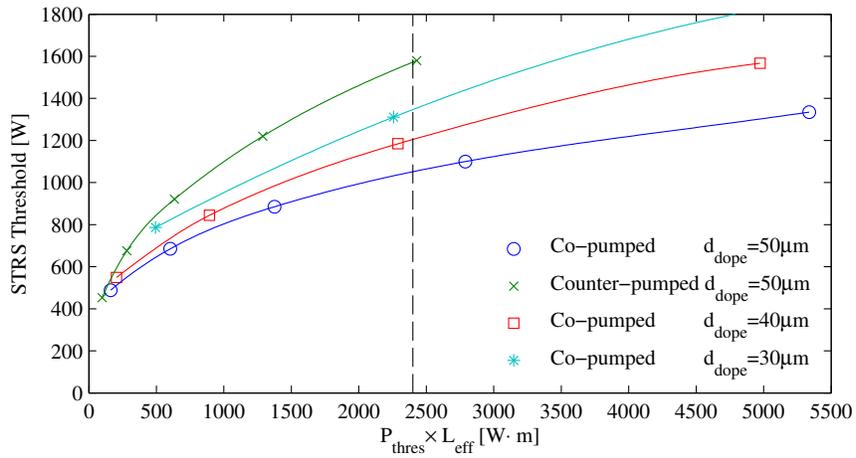}
\caption{\label{fig.PL}Plot of signal threshold output power versus $P_{\rm thres}L_{\rm eff}$ for the 50 $\mu$m core fiber. The curves are spline fits to the computed data points. For $P_{\rm thres}L_{\rm eff}=2400$ W$\cdot$m ($\gamma=6$) the threshold powers are 1050 W for co-pumped and fully doped; 1205 W for co-pumped and $d_{\rm dope}=40$ $\mu$m; 1347 W for co-pumped and $d_{\rm dope}=30$ $\mu$m; 1572 W for counter-pumped and fully doped.}
\end{figure}

In Fig. \ref{fig.PL} we plot the STRS threshold powers versus the quantity $P_{\rm thres}L_{\rm eff}$ for our example fiber with $d_{\rm core}=50$ $\mu$m. Selecting an arbitrary value for $P_{\rm thres}L_{\rm eff}$ on the horizontal axis (or equivalently the value of $\gamma$), the values of the STRS threshold can be read from the computed curves. From the four curves it appears that for any value of $\gamma$ the counter-pumped fully doped amplifier provides the highest STRS threshold power, the co-pumped fully doped amplifier offers the lowest STRS threshold power, while the confined doping, co-pumped amplifiers give intermediate threshold powers.

\section{Scaling for other core/cladding sizes}

To first order changing the core diameter while keeping the ratio $d_{\rm core}/d_{\rm clad}$ fixed, and keeping $NA=0.054$, the threshold powers do not change. The frequency offset changes proportional to $1/A_{\rm eff}$ but the degree of saturation, and thus the value of $\chi^{\prime}$ is unaltered.

However, for $d_{\rm core}<30$ $\mu$m, the value of $V$ is reduced to less than 5, so the mode overlap with the core becomes noticeably weaker, which tends to raise the STRS threshold\cite{dong,hansen}. At the other extreme, for $d_{\rm core}>80$ $\mu$m, thermal lensing becomes significant, and this constriction of the modes may alter the STRS threshold. We will examine these cases more closely in future studies.

As $d_{\rm core}$ changes the area $A_{\rm eff}$ changes and this changes the value of $\gamma$ required to suppress SBS for a given value of $P_{\rm thres}L_{\rm eff}$ according to Eq. \ref{eq.gamma}. Similar powers are possible, while avoiding both SBS and STRS, but the value of $\gamma$ necessary to avoid SBS scales approximately as $1/d_{\rm core}^2$. If there is a constraint on $\gamma$ imposed by beam combining requirements, for example, this implies a lower limit on the core size.

\section{Scaling for other LP$_{11}$ starting powers}

The thresholds computed in this study are based on an input signal power of $10^{-16}$ W in LP$_{11}$, with a frequency shift to the STRS gain maximum, or threshold minimum. This input power level corresponds approximately to the quantum noise limit, and serves as a standard point of comparison among models. Actual starting powers will be considerably higher than this if there is amplitude modulation on the pump or signal seed with a modulation frequency near the optimum STRS frequency shift\cite{smithsmith2012a,smithsmith2013b}. Thermal noise is probably 2-3 factors of 10 higher than this as well. More realistic thresholds would be based on input powers of perhaps 10$^{-10}$ W, but the exact level will require measurements of the modulation properties of the pump and seed. The reference thresholds presented here ($P_{\rm ref}$) can be used to estimate the threshold for a higher input power ($P_{\rm start}$) using
\begin{equation}
P_{\rm thres}=P_{\rm ref}\frac{\log(P_{\rm start}/10)}{\log(10^{-16}/10)}
\end{equation}
where the divisors of 10 are present because our definition of threshold as 1\% of the power in LP$_{11}$ implies a threshold output power of order 10 W in LP$_{11}$.

The thresholds computed in this study are also based on negligible signal loss due to absorbing impurities in the glass or due to photodarkening. The presence of these processes adds heat with a profile that contributes to STRS gain, and so reduces the thresholds\cite{smithsmith2013c}.

\section{Conclusion}

We showed that transverse hole burning, or saturation of the population inversion, in a Yb$^{3+}$ doped fiber can strongly influence the STRS or mode coupling gain, and this effect can be exploited to substantially raise mode instability thresholds. Saturation can also be used in conjunction with confined doping, both contributing to raising the threshold. 

The benefits of high saturation persist when SBS suppression is considered, even though high saturation implies longer fibers with higher $P_{\rm thres}L_{\rm eff}$ values. The benefits of confined doping also survive the requirement of SBS suppression.

Fibers with high saturation have the added advantage of minimizing photodarkening, assuming photodarkening increases with higher upper state population density, as experiments seem to indicate\cite{soderlund}.

\clearpage
\section{Tables}\label{sec.Tables}
In the following tables $\Delta \nu$ is the frequency of LP$_{01}$ minus the frequency of LP$_{11}$.

\begin{table}[htb]
\centering\caption{Thresholds: co-pumped, $d_{\rm core}=50$ $\mu$m, $d_{\rm dope}=50$ $\mu$m }
\begin{tabular}{ccccccc}
$d_{\rm clad}$ [$\mu$m] &$\Delta \nu$ [Hz]&$L$ [m]&$P_{\rm thres}$ [W]& $\int Pdz$ [W$\cdot$m]&$L_{\rm eff}$&$\gamma$\\
\hline
100&1100&0.8&488&161&0.330&1\\
200&1100&1.6&685&603&0.880&1.51\\
300&1100&2.6&885&1375&1.55&3.44\\
400&1100&4.0&1101&2789&2.53&6.97\\
500&1100&6.0&1335&5338&4.00&13.3\\\hline
\end{tabular}
\end{table}
\begin{table}[htb]
\centering\caption{Thresholds: counter-pumped, $d_{\rm core}=50$ $\mu$m, $d_{\rm dope}=50$ $\mu$m }
\begin{tabular}{ccccccc}
$d_{\rm clad}$ [$\mu$m] &$\Delta \nu$ [Hz]&$L$ [m]&$P_{\rm thres}$ [W]& $\int Pdz$ [W$\cdot$m]&$L_{\rm eff}$&$\gamma$\\
\hline
100&1100&0.8&453&98&0.216&1\\
200&1100&1.6&676&281&0.416&1\\
300&1100&2.6&921&634&0.688&1.58\\
400&1100&4.0&1220&1288&1.06&3.22\\
500&1100&6.0&1580&2429&1.54&6.07\\\hline
\end{tabular}
\end{table}
\begin{table}[htb]
\centering\caption{Thresholds: co-pumped, $d_{\rm core}=50$ $\mu$m, $d_{\rm dope}=40$ $\mu$m }
\begin{tabular}{ccccccc}
$d_{\rm clad}$ [$\mu$m] &$\Delta \nu$ [Hz]&$L$ [m]&$P_{\rm thres}$ [W]& $\int Pdz$ [W$\cdot$m]&$L_{\rm eff}$&$\gamma$\\
\hline
100&1400&0.9&549&204&0.372&1\\
200&1400&1.9&844&893&1.06&2.23\\
300&1400&3.2&1185&2290&1.93&5.72\\
400&1400&5.0&1567&4974&3.17&12.4\\\hline
\end{tabular}
\end{table}
\begin{table}[htb]
\centering\caption{Thresholds: co-pumped, $d_{\rm core}=50$ $\mu$m, $d_{\rm dope}=30$ $\mu$m }
\begin{tabular}{ccccccc}
$d_{\rm clad}$ [$\mu$m] &$\Delta \nu$ [Hz]&$L$ [m]&$P_{\rm thres}$ [W]& $\int Pdz$ [W$\cdot$m]&$L_{\rm eff}$&$\gamma$\\
\hline
100&1900&1.4&786&493&0.627&1.23\\
200&1900&3.0&1311&2257&1.72&5.64\\
300&1900&6.0&1975&7757&3.93&19.4\\\hline
\end{tabular}
\end{table}

\end{document}